\documentclass[usenatbib]{mnras}
\usepackage{aas_macros}
\usepackage{natbib}
\usepackage[utf8]{inputenc}
\usepackage{amsfonts}
\usepackage{amsmath}
\usepackage{amssymb}
\usepackage{graphicx}
\usepackage[dvipsnames]{xcolor}
\usepackage{hyperref}
\hypersetup{colorlinks=true,allcolors=teal}
\usepackage{microtype}

\newcommand{\Mh}{\ensuremath{h^{-1}M_{\odot}}}

\newcommand{\Mpch}{\ensuremath{h^{-1}{\rm Mpc}}}
\newcommand{\kpch}{\ensuremath{h^{-1}{\rm kpc}}}

\newcommand{\Mmer}{\ensuremath{\Gamma_{\rm mer}}}
\newcommand{\Mdif}{\ensuremath{\Gamma_{\rm dif}}}
\newcommand{\Mtot}{\ensuremath{\Gamma_{\rm tot}}}

\newcommand{\avg}[1]{\ensuremath{\left\langle \,#1\, \right\rangle}}
\newcommand{\e}[1]{\ensuremath{{\rm e}^{#1}}}

\newcommand{\der}{\ensuremath{{\rm d}}}

\newcommand{\eqn}[1]{equation~\eqref{#1}}

\newcommand{\be}{\begin{equation}}
\newcommand{\ee}{\end{equation}}

\title[Mass accretion rates and environment]{Mass accretion rates and multi-scale halo environment in cold and warm dark matter cosmologies}
\author[Dhoke \& Paranjape]{
Payaswinee Dhoke$^{1}$\thanks{E-mail: payas0906@gmail.com}, 
Aseem Paranjape$^{2}$\thanks{E-mail: aseem@iucaa.in}
\\  
 $^1$ Dharampeth M. P. Deo Memorial Science College, North Ambazari Road, Nagpur 440033, India\\
 $^2$ Inter-University Centre for Astronomy \& Astrophysics, Ganeshkhind, Post Bag 4, Pune 411007, India}
\date{draft}

\begin{document}

\maketitle
\begin{abstract}
We study the evolving environment dependence of mass accretion by dark  haloes in simulations of cold and warm dark matter (CDM and WDM) cosmologies. 
The latter allows us to probe the nature of halo growth at scales below the WDM half-mode mass, which form an extreme regime of nonlinear collisionless dynamics and offer an excellent test-bed for ideas relating to hierarchical growth. 
As environmental proxies, we use the local halo-centric matter density $\delta$ and tidal anisotropy $\alpha$, as well as large-scale halo bias $b_1$. 
Our analysis, while reproducing known trends for  environment-dependent accretion in CDM, as well as the comparison between accretion in CDM and WDM, reveals several interesting new features. 
As expected from excursion set models, WDM haloes have higher specific accretion rates, dominated by the accretion of diffuse mass, as compared to CDM haloes. 
For low-mass WDM haloes, we find that the environment-dependence of both diffuse mass accretion as well as accretion by mergers is almost fully explained by $\alpha$. 
For the other cases, $\delta$ plays at least a comparable role. 
We detect, for the first time, a significant and evolving assembly bias due to diffuse mass accretion for low-mass CDM and WDM  haloes (after excluding splashback objects), with a $z=0$ strength higher than with almost all known secondary variables and largely explained by $\alpha$. 
Our results place constraints on semi-analytical merger tree algorithms, which in turn could affect the predictions of galaxy evolution models based on them. 
\end{abstract}


\label{firstpage}
\pagerange{\pageref{firstpage}--\pageref{lastpage}}

\section{Introduction}

The growth of gravitationally bound haloes of collisionless cold dark matter (CDM) through accretion and mergers is one of the primary physical processes of interest in the hierarchical structure formation paradigm. The stochasticity inherent in the initial cosmological seed fluctuations, coupled with the nonlinearity of gravitational evolution of a collisionless fluid, renders this problem analytically challenging, although considerable insights may be gained through simplified models of structure formation \citep{bcek91,lc93,ms14-markov}. Since dark haloes are expected to be the cradles of \emph{galaxy} formation and evolution \citep{wr78}, understanding the evolving nature of halo mass accretion and its dependence on the local and large-scale environment of haloes is expected to yield important clues into corresponding correlations in the observed distribution and evolution of galaxies in the Universe.

The understanding that accretion rates in the late Universe are sensitive to the shape of the \emph{initial} power spectrum \citep[][see also below]{lc93} suggests a useful tool for investigating the nature of mass accretion, in the form of CDM-like power spectra which are suppressed at small scales (large $k$), mimicking the effects of a thermally produced warm dark matter (WDM) particle. The steep cut-off in power in such models creates an extreme situation where haloes forming close to the corresponding mass scale experience dramatically enhanced growth, and haloes below this scale simply do not exist \citep[e.g.,][]{angulo}. Simulations performed with WDM-like power spectra, therefore, offer ideal test-beds for the environment dependence of mass accretion: any physical model that purports to explain the nature of mass accretion in CDM must also do so for WDM, since the physics of a self-gravitating collisionless fluid is common to both \citep{hp14}. Of course, WDM models are physically interesting in their own right, from the point of view of small-scale challenges for the CDM framework \citep[see][for a review]{bb-k17}, although this is not the focus of the present work.

There has been considerable work to date studying mass accretion by haloes via mergers and smooth (or diffuse) accretion \citep{fm10, genel, benson+13, elahi}, as well as its dependence on halo environment \citep[defined typically in terms of halo-centric dark matter density using different smoothing schemes, see][see also \citealp{mdk15}]{genel,fm09, fm10,maulbetsch+07,bprg17, lee}. The overall understanding that has emerged from these studies, regardless of the exact choice of definition of halo-centric density, is that denser environments tend to promote mergers while underdense environments are more conducive to diffuse accretion. There are also stark differences between accretion in CDM and WDM, particularly at low mass, as expected from the discussion above, with low-mass WDM haloes accreting rapidly and primarily through diffuse accretion.

A related line of study is that of \emph{halo assembly bias} or secondary bias, i.e., the correlation (at fixed halo mass) between secondary halo properties other than mass and halo-centric density (or bias) measured at cosmological scales \citep{st04,gsw05}. Although assembly bias has been studied over a wide range of halo mass and redshift using many choices of secondary variables such as age \citep{st04,gsw05,jsm07}, concentration \citep{wechsler+06,abl08}, shape \citep{fw10, vDaw12}, angular momentum or spin \citep{gw07} and velocity dispersion structure \citep{fw10}, the correlation between halo bias and mass accretion rate has thus far been limited to massive objects at $z=0$ \citep{lazeyras}. Local halo environment has been found to influence halo assembly bias of various secondary variables \citep{Wang07, Jung14, lee, Yang17, Musso18}. Moreover, recent work on assembly bias at $z=0$ has revealed the importance of the \emph{local tidal environment} of haloes \citep{hpdc09, bprg17,phs18a} 
in explaining the assembly bias of many secondary variables, including concentration, spin, shape and velocity dispersion structure over a wide range of halo mass \citep[][see also \citealp{han+19}]{rphs19}. It is therefore very interesting to ask whether the halo tidal environment plays a similar role in explaining any assembly bias trends with mass accretion.

In this work, we perform a detailed study of the nature of the environment dependence of mass accretion by haloes, segregated into contributions due to mergers and diffuse mass, as a function of redshift and for a range of halo masses. We will do so using $N$-body simulations of both CDM and WDM cosmologies; as mentioned above, the latter will allow us to better resolve the multi-scale environment dependence of mass accretion due to its sensitivity to the shape of the initial matter power spectrum. We will connect these results to the assembly bias literature by performing, for the first time, a comprehensive study of assembly bias due to mass accretion across cosmic time, along with its connection to the evolving local halo environment, for both CDM and WDM. Our results reproduce previously observed trends while extending these to new local environmental variables such as the tidal anisotropy, and are expected to place useful constraints on semi-analytical models of halo formation and growth \citep[e.g.,][]{sk99,pch08,jvdb14}, which in turn form the bedrock of several semi-analytical models of galaxy formation and evolution \citep[e.g.,][]{bb10,benson12,barausse12,dfdp14,birrer+14,spt15,yung+19}.

The paper is organised as follows. We describe  our simulations and analysis techniques in section~\ref{sec:sims}. We present our results along with a discussion in section~\ref{sec:results}, with section~\ref{subsec:mergvsdiff} focusing on environment-independent evolution, sections~\ref{subsec:medianenv} and~\ref{subsec:mdot<->env} focusing on trends with local environment and section~\ref{subsec:assemblybias} devoted to assembly bias. We discuss some of these trends in the framework of the  excursion set approach in section~\ref{sec:analytical} and conclude with a summary of our main results in
section~\ref{sec:conclude}. Throughout, the base-10 logarithm is denoted by  `log' and the natural logarithm by `ln'.

\section{Simulations and Techniques}
\label{sec:sims}
Here we describe our simulations and analysis tools for identifying haloes and measuring their accretion rates and local as well as large-scale environments.

\subsection{Simulations}
We have used $N$-body simulations of CDM and WDM performed using the tree-PM code \textsc{gadget-2} \citep{springel}\footnote{\href{http://www.mpa-garching.mpg.de/gadget/}{http://www.mpa-garching.mpg.de/gadget/}} in a cubic, periodic box of comoving length $150\Mpch$ sampled with $1024^3$ particles, corresponding to a particle mass $m_{\rm  p} = 2.4\times 10^8 \Mh$, with a $2048^3$ PM grid and comoving force resolution $4.9 \kpch$ corresponding to $1/30$ of the Lagrangian inter-particle spacing. 

The CDM transfer function $T_{\rm cdm}(k)$ for generating initial conditions was computed using the code \textsc{camb} \citep{camb}\footnote{\href{http://camb.info}{http://camb.info}} with a spatially flat $\Lambda$CDM cosmology having total matter density parameter $\Omega_{\rm m}=0.276$, baryonic matter density $\Omega_{\rm b}=0.045$, Hubble constant $H_0=100h\,{\rm kms}^{-1}{\rm Mpc}^{-1}$ with $h=0.7$, primordial scalar spectral index $n_{\rm s}=0.961$ and r.m.s. linear fluctuations in spheres of radius $8\Mpch$, $\sigma_8=0.811$, consistent with the 7-year results of the \emph{Wilkinson Microwave Anisotropy Probe} \citep[WMAP7,][]{Komatsu2010}.

For the WDM model, we additionally suppress small-scale power in the linear transfer function as appropriate for the free-streaming of a thermally produced WDM particle with mass $m_{\rm dm}=0.4\,{\rm keV}$ according to the fitting function of \citet{Bode2001} \citep[with parameters taken from][]{Viel2005}
\begin{equation}
 T_{\rm wdm}(k) = T_{\rm cdm}(k) \left[1 + (\alpha k)^{2\mu}\right]^{-5/\mu},
\label{eq:Tk-wdm}
\end{equation}
with $\mu=1.12$ and
\begin{equation}
\alpha  \equiv 0.049 \left(\frac{\Omega_{\rm m}}{0.25}\right)^{0.11} \left(\frac{h}{0.7}\right)^{1.22}
\left( \frac{m_{\rm dm}}{1\,{\rm keV}}  \right)^{-1.11}\,\Mpch\,.
\label{alpha}
\end{equation}
The resulting ``half-mode'' mass-scale \citep[c.f., e.g.,][]{Schneider2012} of $M_{\rm hm}\simeq 3\times10^{11}\Mh$ is resolved with $\sim1200$ particles (see below).\footnote{The collisionless WDM fluid is assumed to be in the perfectly cold limit, i.e., we ignore the small thermal velocity dispersion of a real WDM fluid. This is expected to be accurate at the epochs of our interest, well after perturbations have been suppressed below the largest free-streaming scale in linear theory \citep{angulo}.} Although a WDM particle with $m_{\rm dm}=0.4\,{\rm keV}$ is completely ruled out by Lyman-alpha forest observations as being the dominant component of dark matter \citep{Viel2013,irsic+17,pd+20}, it allows us to resolve the entire initial power spectrum up to the truncation scale with sufficient particles, thus providing a useful test-bed for studying mass accretion and its environment dependence in the absence of small-scale perturbations.

Initial conditions for both CDM and WDM were generated at $z=99$ with the \emph{same random seed} using $2^{\rm nd}$ order Lagrangian perturbation theory \citep{scoccimarro98} with the code \textsc{music}  \citep{hahn11-music}.\footnote{\href{https://www-n.oca.eu/ohahn/MUSIC/}{https://www-n.oca.eu/ohahn/MUSIC/}} Snapshots were stored starting from $z=12$ to $z=0$ at time intervals equally spaced in the scale factor with $\Delta a = 0.004615$ (leading to 201 snapshots), which provides sufficient time resolution for the accretion and merger analysis described below. The simulations and analysis were performed on the Perseus and Pegasus clusters at IUCAA.\footnote{\href{http://hpc.iucaa.in}{http://hpc.iucaa.in}}

\subsection{Halo identification and masses}
\label{subsec:massfn}
Haloes were identified using the code \textsc{rockstar} \citep{behroozi-a}\footnote{\href{https://bitbucket.org/gfcstanford/rockstar}{https://bitbucket.org/gfcstanford/rockstar}} which implements a Friends-of-Friends algorithm in 6-dimensional phase space and provides information on gravitationally bound haloes as well as their substructure. Merger trees were constructed using the 201 snapshots in each simulation (CDM and WDM) using the code \textsc{consistent-trees}  \citep{behroozi-b}.\footnote{\href{https://bitbucket.org/pbehroozi/consistent-trees}{https://bitbucket.org/pbehroozi/consistent-trees}} Since we wish to focus on the accretion rates of well-resolved haloes in this work, we exclude all subhaloes and further consider only those objects for which the virial energy ratio $\eta = 2T/|U|$ satisfies $0.5<\eta<1.5$, which mitigates the effects of unrelaxed objects and numerical artefacts \citep{bett07}. We exclude $\sim 2\%(6\%)$ objects of low mass halo population which do not satisfy this virial ratio condition in CDM(WDM). For the high mass bin these objects form $\sim 4\%$ of the halo population in either cosmology.

We further exclude splashback objects, which spatially mimic genuine haloes at the redshift of interest but have passed through  a larger host in the past and hence are physically closer to subhaloes. We do so using the value of the `first accretion scale' reported by \textsc{consistent-trees}, which records the earliest epoch at which the main progenitor of a given descendant passed through the virial radius of a larger object; we exclude all descendants whose first accretion scale is earlier than the epoch of interest.\footnote{To avoid effects of numerical precision in comparing scale factor values, we define splashbacks as objects whose first accretion epoch is more than 100 Myr in the past of the epoch of interest. We have checked that varying this threshold by factor 2 on either side makes no difference to our results.} For completeness, we report that splashback haloes at $z=0$ form $\lesssim 6\% \,(4.5\%)$ by number of our low-mass halo populations in CDM (WDM) and $\lesssim0.7\%$ of high-mass objects in either cosmology (see below for the corresponding mass ranges), with the fractions becoming vanishingly small at high redshifts. With all the selection criteria, we report that, we select, $\sim 80\% (\sim 75\%)$ of low mass ROCKSTAR objects in CDM(WDM) and $\sim 87 \%$  of high mass objects in the either cosmology. We have checked that the selected low mass haloes in CDM and WDM are \emph{relaxed} by plotting the distributions of the positions and velocities of the particles forming haloes. The distributions are found to show single well defined peaks.

Throughout, we will quote halo and progenitor masses using the mass definition $M_{\rm vir}$ as reported by \textsc{rockstar}. This corresponds to the bound mass contained in a halo-centric sphere of virial radius $R_{\rm vir}$ which encloses a density equal to $\Delta_{\rm vir}$ times the critical density of the Universe, where $\Delta_{\rm vir}$ is the spherical collapse overdensity and is taken from the fitting function provided by \citet{bn98} (at $z=0$, $\Delta_{\rm vir}\simeq98$ for our cosmology, while $\Delta_{\rm vir}\to18\pi^2\simeq178$ at high-$z$ during matter domination). Additionally, we use the \textsc{rockstar} values of $M_{\rm 200b}$ to infer the halo-centric radius $R_{\rm 200b}$ which encloses a density equal to 200 times the mean density of the Universe at each redshift; this will be useful when characterising local halo environments below.

\begin{figure}
\centering
\includegraphics[width=0.485\textwidth,trim= 2 4 20 20,clip]{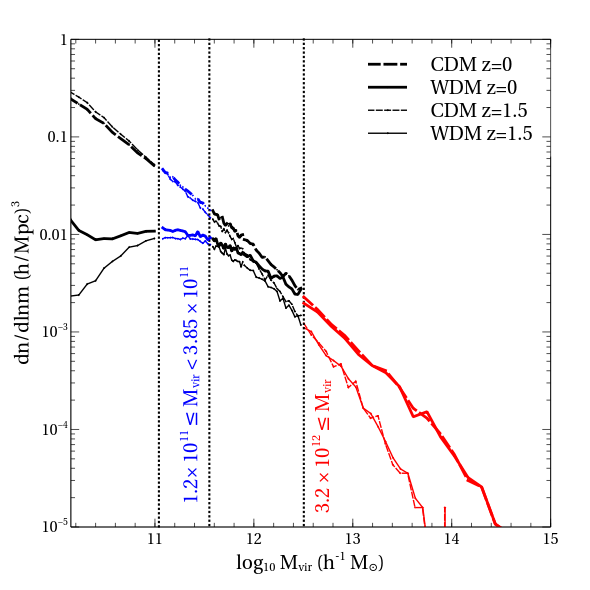}
\caption{{\bf Halo mass function:} Dashed (solid) lines show the halo mass function in CDM (WDM) at $z=0.0$ (thick curves) and $z=1.5$ (thin curves) as a function of virial mass $M_{\rm vir}$. The horizontal axis starts at $M_{\rm vir}= {1.4\times10^{10}\Mh}$, corresponding to a 60 particle cut which we use for identifying progenitors. The first vertical dotted line from the left indicates $M_{\rm vir}=1.2\times10^{11}\Mh$ or 500 particles, which we use as the threshold for identifying descendants at any redshift (see text for other criteria applied in selecting a clean halo sample). The low-mass and high-mass bins we use are indicated in blue and red colours, respectively, demarcated by the remaining vertical lines and with the corresponding ranges mentioned in the labels.
\underline{\emph{Highlights:}} In the WDM mass function, we clearly see the suppression of halo counts below the half-mode mass $M_{\rm hm}\simeq3\times10^{11}\Mh$, and (at low redshift) the spurious upturn at very low masses due to numerical artefacts. 
} 
\label{massfunction}
\end{figure}

Figure \ref{massfunction} shows the mass functions for haloes in CDM (dashed) and WDM (solid) at $z=0$ and $z=1.5$. The vertical lines demarcate the mass bins we will use below; the lowest mass threshold corresponds to 500 particles and the mass functions in the low-mass (high-mass) bin we employ are coloured blue (red). 
The CDM mass function shows the well-studied power law rise towards low masses and exponential decline at high masses \citep{ps74,bcek91,st99,Tinker08}. 

The WDM mass function is identical to the CDM one at high masses for each redshift, but turns over around the half-mode mass (see above), leading to a suppression of halo abundances at low masses as expected from the lack of small-scale perturbations (\citealp{hp14}; see also \citealp{benson+13,ssr13}). At masses smaller than $\sim10^{10.4}\Mh$, however, we see an \emph{up-turn} in the $z=0$ WDM mass function, which is a well-known consequence of numerical artefacts in the $N$-body technique applied to initial conditions with suppressed small-scale power \citep{W&W,angulo,Lovell2014,A&C}. Note that the virial cut of $0.5<\eta<1.5$ mentioned earlier already removes many spurious objects at masses close to the half-mode mass \citep{A&C}; removing the lower-mass spurious objects would require a more sophisticated study of the \emph{initial} proto-patches from which these objects evolve \citep{Lovell2014}, which we have not performed here.

Our choice of the lowest mass bin for \emph{descendant} haloes is sufficiently far above the mass scale where spurious objects start becoming numerically relevant. We have checked this by applying the correction suggested by \citet{ssr13} by fitting a power law to the mass function below the up-turn scale (not displayed) and subtracting it from the measured mass function; the result agrees with the measured mass function in the mass bins of interest at better than $\sim2\%$ at all redshifts. Spurious objects therefore do not contribute to our chosen populations of descendant haloes at any redshift. However, when identifying \emph{progenitor} haloes in the merger trees, we will employ a lower mass cut of 60 particles (the left edge of the horizontal axis in figure~\ref{massfunction}). This affects the quantification of mass accretion rates in the WDM case at low redshifts, by artificially enhancing the accretion due to mergers and correspondingly decreasing the accretion of diffuse mass. We return to this point below.

\subsection{Measuring specific mass accretion rates}
\label{subsec:Mdot-measure}
For the analysis below, we use haloes in the two mass bins discussed above, all of which are well-resolved with more than 500 particles.
We used merger trees generated by \textsc{consistent-trees} to find all the progenitors of a given halo within the previous 2 dynamical times\footnote{We have repeated our main analysis for accretion rates calculated over 1 dynamical time and find qualitatively similar but noisier results. We therefore focus on results using accretion rates over 2 dynamical times.} at any chosen redshift. The values of the dynamical time $T_{\rm dyn}$ and the redshift interval $\Delta z\equiv z_{\rm i}-z_{\rm f}$ corresponding to $2T_{\rm dyn}$ at different redshifts are given in table \ref{tab:table1}. Our snapshot resolution of $\Delta a\simeq0.005$ (see above) provides us with excellent sampling of the required $\Delta z$ at all the redshifts we probe (see table~\ref{tab:table1}).

\begin{figure*}
\centering
\includegraphics[width=0.9\textwidth,trim= 2 0 20 20,clip]{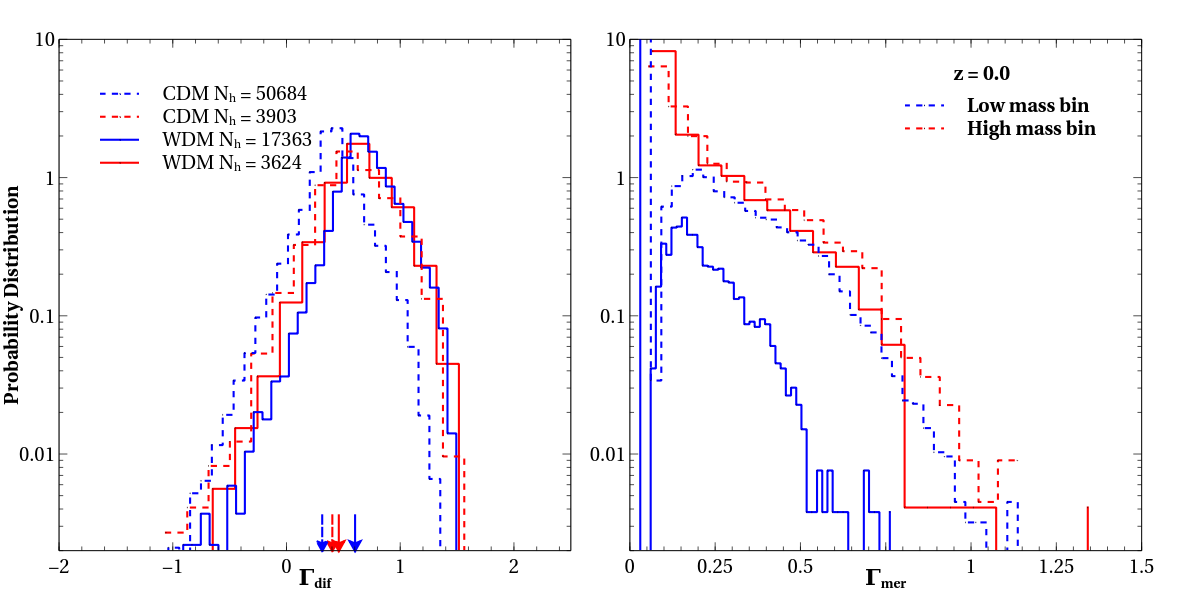}
\caption{
{\bf Accretion rate distributions:}
Histograms in the \emph{left (right) panel} show the distributions of the specific mass accretion rates \Mdif\ (\Mmer) due to diffuse mass (mergers) at $z=0$ (see equations~\ref{eq:Mdot-def}). Dashed (solid) lines show results for CDM (WDM) haloes, with the red (blue) curves showing results for the high-mass (low-mass) bin. The legend in the \emph{left panel} indicates the number of haloes that contributed to each mass bin, while the arrows mark the median values of \Mdif\ for the respective distributions. \underline{\emph{Highlights:}} Low-mass WDM haloes have the highest \Mdif\ and lowest \Mmer\ values on average. 
} 
\label{Mrates-distri}
\end{figure*}

Throughout, we retain progenitors that are resolved with more than 60 particles ($M_{\rm vir}>1.4 \times 10^{10}\Mh$).  
This leads to a different dynamic range in progenitor-to-descendant mass ratio $\chi$ for the low-mass ($\chi\gtrsim0.04$) and high-mass bin ($\chi\gtrsim6\times10^{-5}$). To the extent that accretion rates are self-similar with mass \citep{fg84,Bertschinger85b}, this would lead to accretion by `mergers' being systematically enhanced in high-mass haloes simply because relatively low-mass objects are counted as progenitors as compared to the low-mass bin.
Provided accretion rates are accurately estimated without double-counting due to (artificial) fragmentation, however, the contribution to genuine mergers from very low mass ratios (say $\chi<10^{-2}$) should be small \citep{genel}. 
\textsc{consistent-trees} improves the consistency of progenitor/descendant assignment across time steps by solving differential equations for the expected locations of potential descendants \citep{behroozi-b}, so that effects of artificial fragmentation are reduced.
Also, since we define accretion rates over $2T_{\rm dyn}$, any residual effects of instantaneous fragmentation should typically be averaged over. 
Moreover, environmental trends have been found to be relatively insensitive to whether progenitors are defined using a fixed minimum particle count or minimum $\chi$ \citep{fm10}.
We therefore proceed with our analysis using a fixed threshold of 60 particles on progenitor mass and comment below on analyses using other choices. 

We follow \citet{fm10} and divide the mass accreted by a halo into two parts: (i) via other resolved haloes as mergers and (ii) in the form of diffuse mass, which includes bound structures below the chosen mass resolution, actually unbound particles which may have been tidally stripped from other objects, or genuinely diffuse mass which has never been part of a bound structure earlier.
In practice, we calculate the mass accreted in a given time interval $z_{\rm i} > z > z_{\rm f}$ due to mergers as the total mass of all progenitors except the main progenitor. Correspondingly, diffuse mass accreted in the same time interval is the difference between the descendant halo's mass and the total mass of all the progenitors. Finally, the total mass accreted in this time interval is the sum of the masses accreted via mergers and diffuse accretion, i.e., the difference between the descendant halo's mass and the mass of the main progenitor. Symbolically, if we define

$M_0$ : virial mass of descendant halo at $z=z_{\rm f}$,

$M_j$ : virial mass of the $j^{th}$ progenitor halo at $z=z_{\rm i}$,

$M_1$ : virial mass of the main
progenitor at $z=z_{\rm i}$,

\noindent
then the \emph{dimensionless specific accretion rates} \Mmer\ (merger), \Mdif\ (diffuse) and \Mtot\ (total) -- where $\Gamma\sim\der\ln M/\der\ln a$ \citep[e.g.,][]{dk14} -- can be written as 
\begin{align}
    \Mmer &\equiv \frac{\left(\sum_{j\ge2}M_j\right)}{M_0 (z_{\rm i}-z_{\rm f})}\times(1+z_{\rm f})\,, \notag\\
    \Mdif &\equiv \frac{\left(M_0 - \sum_{j\ge1}M_j\right)}{M_0 (z_{\rm i}-z_{\rm f})}\times(1+z_{\rm f})\,,  \notag\\
    \Mtot &\equiv \frac{\left(M_0 - M_1\right)}{M_0 (z_{\rm i}-z_{\rm f})}\times(1+z_{\rm f}) \,.
    \label{eq:Mdot-def}
\end{align}
We have chosen a convention such that positive values of the rates correspond to increase in mass from $z_{\rm i}$ to $z_{\rm f}$.

\begin{table}
\centering
    \caption{Value of dynamical time $T_{\rm dyn}$ in Gyr, the redshift interval $\Delta z$ corresponding to $2\,T_{\rm dyn}$ in the past, and the corresponding number of simulation snapshots $N_{\rm snap}$ tracked, at each redshift studied in this work.}
    \label{tab:table1}
    \begin{tabular}{|c | c | c | c | c | c | c | c |} 
    \hline\hline
    $z$ & 0.0 & 0.25 & 0.5 & 0.8 & 1.0 & 1.5 & 2.0  \\
    \hline\hline
    $T_{\rm dyn}$ (Gyr) & 3.14 & 2.55 & 2.08 & 1.67 & 1.46 & 1.08 & 0.83\\
    \hline
    $\Delta z\,(2 T_{\rm dyn})$ & 0.68 & 0.80 & 0.95 & 1.11 & 1.20 & 1.50 & 2.23 \\
    \hline
    $N_{\rm snap}$ & 89 & 70 & 57 & 47 & 42 & 34 & 32\\
    \hline\hline
    \end{tabular}
\end{table}

Using the progenitors selected as above, we computed \Mmer, \Mdif\ and \Mtot\ within the previous 2 dynamical times at seven redshifts $z=0.0$, 0.25, 0.5, 0.8, 1.0, 1.5 and 2.0 for the CDM and WDM simulations.\footnote{As a check, we also explicitly calculated \Mtot\ over $100$ My, finding that these are within $\sim10\%$ of the values reported by \textsc{consistent-trees}. The differences are likely due to different choices of progenitor mass resolution between our values and the default settings of \textsc{consistent-trees}, and are not expected to alter our conclusions qualitatively.}
These rates are found to show distributions similar to those shown in \citet{fm10}. Figure \ref{Mrates-distri} shows the normalised probability distributions for $\Mdif$ \emph{(left panel)} and $\Mmer$ \emph{(right panel)} for $z=0$ haloes in the low-mass (blue) and high-mass (red) bin, for CDM (dashed lines) and WDM (solid lines). Arrows in the left panel (with identical colour-coding and line styles) indicate the corresponding median values of $\Mdif$.  These arrows show that the median \Mdif\ is the highest for WDM low-mass haloes. For the high-mass bin, CDM and WDM haloes show very similar distributions of the two accretion rates. $\Mmer$ is always positive except if there were no recent mergers, in which case $\Mmer=0$ exactly. $\Mdif$ can be negative if the total mass of progenitors is more than the mass of the descendant halo, which can happen due to tidal stripping; this is evidently a significant effect only for low-mass CDM haloes. Below, we will investigate in detail the redshift evolution and environmental trends of both \Mdif\ and \Mmer.

As mentioned earlier, the dominant presence of spurious objects in the WDM simulation at low redshifts and masses $\lesssim10^{10.4}\Mh$ affects the measurement of WDM accretion rates. In particular, \Mmer\ is overestimated and \Mdif\ correspondingly underestimated due to the fact that mass is locked up in spuriously identified objects. However, the \emph{amount} of mass in these low-mass spurious objects is not dramatically large. 
By integrating the measured and power-law-corrected (see section~\ref{subsec:massfn}) mass functions, weighted by mass, over the range $60\,m_{\rm p} \leq M_{\rm vir} \leq 500\,m_{\rm p}$, we find that the contribution of spurious objects in this range is $\simeq40\%$ by mass at $z=0$ and falls to $\lesssim4\%$ by $z=1.5$. Moreover, we will see below that accretion due to mergers is subdominant in WDM in any case. We therefore tentatively conclude that, although the WDM merger accretion rates are likely to be systematically overestimated due to the presence of spurious objects, accounting for this spurious contribution is not expected to alter any of our qualitative conclusions.

\begin{figure*}
\centering
\includegraphics[width=0.9\textwidth,trim= 2 0 20 20,clip]{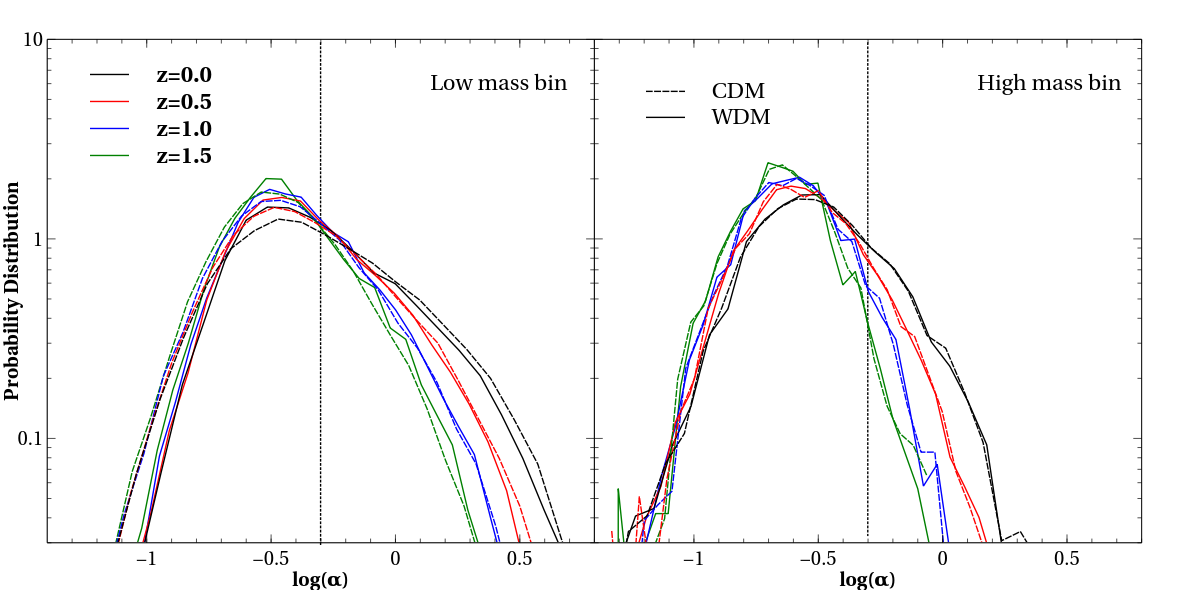}
\caption{{\bf Local tidal anisotropy:} Normalised distributions of the tidal anisotropy $\alpha$ (equation~\ref{eq:alpha-def}) for low-mass \emph{(left panel)} and high-mass \emph{(right panel)} haloes in CDM (dashed curves) and WDM (solid curves) at four redshifts between $0\leq z\leq1.5$ (indicated by colours). Vertical line in each panel indicates $\alpha=0.5$, with higher (lower) values corresponding to filamentary (node-like) environments. 
\underline{\emph{Highlights:}} Distributions for low-mass WDM haloes are systematically narrower than their CDM counterparts.}
\label{cdmwdm-alpha-4z}
\end{figure*}

To assess the effects of the potential caveats discussed above, namely, the effect of a fixed particle threshold versus a threshold in mass ratio for defining progenitors, and the effect of spurious  WDM haloes, we have repeated our entire analysis using four different selection criteria.  In the first modification, we replaced the fixed cut of $>60$ particles per progenitor with a lower threshold of $>30$ particles, which should emphasize the effects of spurious WDM haloes. In the second case, we repeated the analysis for low mass WDM only, with $>100$ particle progenitors, which should completely remove the contribution from spurious WDM haloes. In the third, we used $>60$ particles and imposed relaxation criteria $0.5 < \eta < 1.5$ (which was not done in the default analysis) on the progenitors at $2T_{dyn}$. In the fourth, we used the 60 particle  threshold for progenitors of the low-mass descendants, but a threshold of $M_{\rm vir} > 4.84\times10^{11}\Mh$ for progenitors of high-mass descendants. This corresponds to $\chi\simeq0.08$  with respect to the median descendant mass in the high-mass bin, which is the same as the ratio between the 60 particle threshold and the median mass of the low-mass bin, and should emphasize any differences between environmental trends due to the different progenitor selection. It has been studied that \citep{genel, WangJ11} with changing the mass ratio threshold of most massive progenitor to mergers, the merger accretion rates are found to change significantly. We find that \emph{all our qualitative results remain unchanged in each case}, with only a few minor differences which we comment on later. This indicates that (a) the environmental trends we study are indeed largely insensitive to the choice of progenitor threshold mass and (b) spurious haloes in WDM \emph{do not affect} the inference of the trends. All our results below will be quoted for the (more conservative) fixed threshold of $>60$ particles per progenitor (without imposing $0.5 < \eta < 1.5$ criteria) for each mass bin.

\subsection{Quantifying local halo environment}

We quantify the \emph{local} environment of haloes using scalars constructed from the smoothed halo-centric tidal tensor $T_{ij}(x) = \partial_i \partial_j \Psi_R (x)$, where the Newtonian potential $\Psi_R$ satisfies the normalised Poisson equation $\nabla^2\Psi_R = \delta_R$ with $\delta_R$ being the halo-centric dark matter overdensity Gaussian-smoothed on comoving scale $R$. 
In practice, for any smoothing scale $R$, we invert the Poisson equation in Fourier space using cloud-in-cell (CIC) interpolation for the unsmoothed density on a $512^3$ grid. Namely, we Fourier transform the CIC overdensity to obtain $\delta(\mathbf{k})$, using which the tidal tensor is the inverse Fourier transform of $(k_ik_j/k^2)\delta(\mathbf{k})\e{-k^2R^2/2}$.

As the two scalars of choice, we use the overdensity  $\delta_R$ itself, and the tidal anisotropy $\alpha_R$ introduced by \citet{phs18a}.
If the eigenvalues of $T_{ij}$ are denoted $\lambda_1 \leq \lambda_2 \leq \lambda_3$, then we have
\be
\delta_R = \lambda_1 + \lambda_2 + \lambda_3 
\label{eq:delta-def}
\ee
and
\be
\alpha_R = \sqrt{q_{R}^2} / (1+\delta_R)
\label{eq:alpha-def}
\ee
where $q_{R}^2$ is the tidal shear \citep{heavens, catelan} defined as,
\be
q_{R}^2 = \frac{1}{2} [(\lambda_3 - \lambda_1)^2 + (\lambda_3 - \lambda_2)^2 + (\lambda_2 - \lambda_1)^2]\,.
\ee
For the reasons discussed by \citet{phs18a} and \citet{rphs19}, we define the  local halo environment at scales $R = 4R_{\rm 200b}/\sqrt{5}$ for each halo. (In practice, we first evaluate the tidal tensor on the grid for a fixed set of smoothing scales and then interpolate to the adaptive scale and spatial position corresponding to each halo.) This adaptive filtering choice maximises the correlation between the local halo tidal environment and large-scale halo bias, useful for studies of assembly bias. Hereafter, for brevity we will write $\alpha\equiv\alpha_{R}$ and  $\delta\equiv\delta_{R}$.

A common approach to classifying cosmic web environment is by counting the signs of the eigenvalues of the tidal tensor or density Hessian defined at some fixed smoothing scale; with the tidal tensor we would have \citep[e.g.,][]{hpcd07} $\lambda_1 > 0$ : node, $\lambda_1 < 0$ and $\lambda_2 > 0$ : filament, $\lambda_2 < 0$ and $\lambda_3 > 0$ : sheet, $\lambda_3 < 0$ : void.
The use of $\alpha$ and $\delta$ provides an alternative, continuous definition of halo environment adapted to the halo size. These variables, although correlated, are not completely degenerate \citep[][see also below]{rphs19}. Such a continuous measure of environment is much better suited for studies of large-scale environmental correlations or assembly bias \citep{phs18a,rphs19} as well as environmental trends in galaxy evolution \citep{zphp20}. 

\begin{figure*}
\centering
\includegraphics[width=0.9\textwidth,trim= 2 0 16 20,clip]{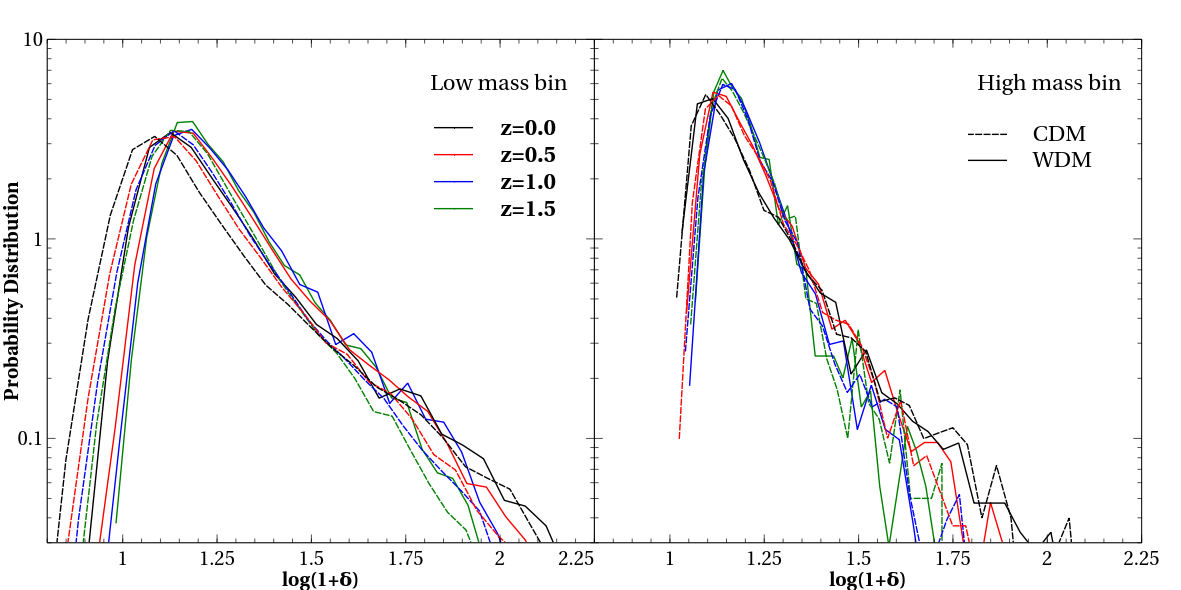}
\caption{{\bf Local overdensity:} Same as figure~\ref{cdmwdm-alpha-4z}, showing results for halo-centric local overdensity $\delta$ (equation~\ref{eq:delta-def}).
\underline{\emph{Highlights:}} As with $\alpha$, the distributions of $\delta$ for low-mass WDM haloes are systematically narrower than their CDM counterparts, and also have higher median values.}
\label{cdmwdm-delta-4z}
\end{figure*}

Physically, small values of the adaptively defined $\alpha$ correspond to isotropic environments (nodes), while large values correspond to very anisotropic environments (filaments). As discussed by \citet{phs18a} and \citet{paranjape20}, the value $\alpha=0.5$ forms a rather sharp threshold separating haloes in node-like and filament-like environments. We have checked that this is true at all the redshifts we study in this work. 

Figure \ref{cdmwdm-alpha-4z} shows the distribution of $\alpha$ at four redshifts for CDM (dashed) and WDM (solid) haloes in the low-mass \emph{(left panel)} and high-mass bin \emph{(right panel)}. The distribution of $\alpha$ at fixed redshift depends on halo mass, with massive haloes having preferentially lower values of $\alpha$ and low-mass haloes spanning a wide range of $\alpha$. Haloes at higher redshifts also have preferentially lower values (as well as narrower distributions) of $\alpha$, which is possibly mainly a consequence of mass resolution (since haloes at fixed mass are rarer in the past), but might also be partially reflecting a genuine evolution of the local cosmic web environment of haloes. 
Figure~\ref{cdmwdm-delta-4z} is formatted identically to figure~\ref{cdmwdm-alpha-4z} and shows the corresponding distributions of $\delta$. At fixed redshift, we see that the distributions for low-mass haloes have wider tails than for high-mass haloes, at both low and high density. At fixed mass, high-redshift haloes span substantially narrower ranges of $\delta$ than those at lower redshift.
For both $\alpha$ and $\delta$, the results for WDM and CDM at each redshift are very similar to each other for high-mass haloes, while the distributions for low-mass WDM haloes are narrower than their CDM counterparts (with the median $\delta$ also being systematically higher for WDM). This extends the results of \citet{phs18a}, who studied CDM haloes at $z=0$, to significantly higher redshift as well as WDM cosmologies.

\subsection{Large-scale environment: halo bias}
\label{subsec:bias}
As an indicator of the \emph{large-scale} halo environment, we estimate the linear bias $b_1$ for each halo using the technique outlined by \citet{phs18a} and \citet{pa20}. This is essentially an object-by-object version of the usual cross-correlation definition of bias in Fourier space -- $P_{\rm hm}(k)/P_{\rm mm}(k)$ -- averaged over low-$k$ modes using the weights discussed by \citet{pa20}, where $P_{\rm hm}$ and $P_{\rm mm}$ are the halo-matter cross power spectrum and matter auto-power spectrum, respectively. 

The box size of $150\Mpch$ leads to small-volume systematic effects in the absolute value of $b_1$ measured for each object, due to missing long-wavelength modes. Here, however, we are only interested in the correlation between $b_1$ and other halo properties like accretion rates and halo environment. As demonstrated by \citet{rphs19}, these correlations are relatively insensitive to volume effects and we therefore expect our results to be robust to such systematics.

\section{Results and discussion}
\label{sec:results}
We now turn to our main results, starting with the environment-independent evolution of mass accretion, followed by trends with local environment and finally assembly bias.

\begin{figure}
\centering
\includegraphics[width=0.45\textwidth,trim= 2 0 20 20,clip]{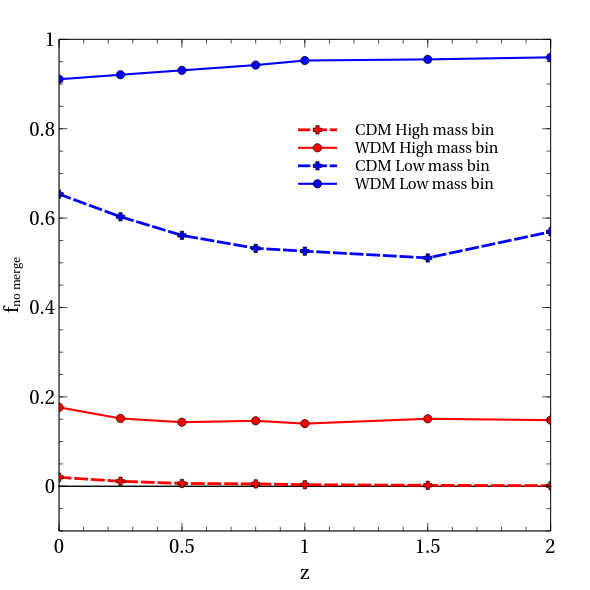}
\caption{{\bf Mergers versus diffuse accretion:}
Evolution with redshift of the fraction $f_{\rm no\,merge}$ of haloes with zero merger rate in the previous 2 dynamical times. Results for the low-mass (high-mass) bin are shown in blue (red), with the dashed (solid) curves showing results for CDM (WDM). \underline{\emph{Highlights:}} Low-mass WDM haloes accrete mainly diffuse mass at all epochs, while essentially every high-mass CDM halo has accreted some mass through mergers in the previous 2 dynamical times, except at very low redshift. } 
\label{fraction}
\end{figure}

\subsection{Mass accretion: mergers vs. diffuse accretion}
\label{subsec:mergvsdiff}

\begin{figure*}
\centering
\includegraphics[width=\textwidth,trim= 2 0 20 20,clip]{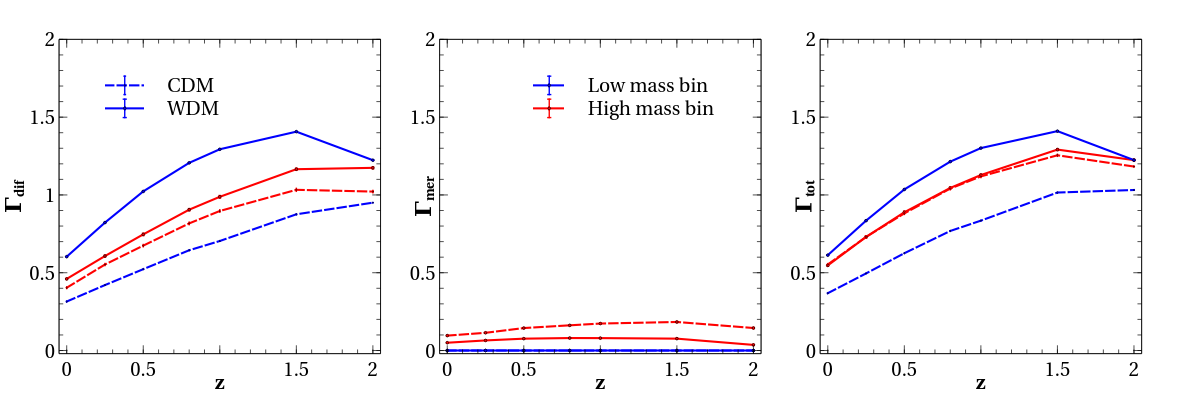}
\caption{{\bf Accretion history:} Evolution with redshift of the median values of the specific accretion rates \Mdif\ \emph{(left panel)}, \Mmer \emph{(middle panel)} and \Mtot\ \emph{(right panel)}, for low-mass (blue) and high-mass (red) haloes in CDM (dashed) and WDM (solid). Error bars were  calculated by bootstrap sampling at each redshift.
\underline{\emph{Highlights:}} \Mdif\ for low-mass WDM haloes is the highest of all categories. In the high-mass bin, although \Mtot\ is nearly identical between CDM and WDM, the relative contribution to this total from mergers versus diffuse accretion is higher in CDM than in WDM. 
} 
\label{MAH}
\end{figure*}

Figure \ref{fraction} shows the evolving fraction $f_{\rm no\,merge}$ of haloes which do not accrete any mass through mergers in the last 2 dynamical times (i.e., haloes without any secondary progenitor larger than the 60 particle threshold, having $\Mmer=0$), for CDM and WDM. High-mass CDM haloes essentially always had mergers, except at very low redshifts where a small fraction have evolved with $\Mmer=0$. At fixed mass, the value of $f_{\rm no\,merge}$ is always larger in WDM than in CDM. This can be easily understood as being due to the lack of small-scale structure and hence fewer low-mass bound structures in WDM. In fact, we see that haloes in the low-mass WDM bin accrete mainly diffuse mass at all redshifts, with $f_{\rm no\,merge}\gtrsim0.9$ at all $z$. This is consistent with corresponding results from \citet{elahi}.     

Figure \ref{MAH} shows the evolution of the median \Mdif\ and \Mmer\ (\emph{left} and \emph{middle panels}, respectively) with the total accretion rate \Mtot\ shown in the \emph{right panel}.\footnote{The trend in the evolution is observed to change near $z\sim1.5$. This trend at high redshift could not be verified with the analytical model and deserves further study.} Error bars on the measurements were calculated by bootstrap sampling: at each redshift and for each mass bin, the accretion rate data are sampled with repetition a number of times and the standard deviation in the median values of each sample gives the value of error.  
We see that WDM high-mass haloes have nearly the same \emph{total} accretion rate as CDM high-mass haloes, at all epochs \citep[consistent with the earlier results by][]{knebe02, benson+13}. WDM low-mass haloes, on the other hand, have higher total accretion rates than, both, CDM low-mass haloes and also WDM  (and CDM) high-mass haloes.

Upon splitting the accretion rate between mergers and diffuse accretion, we see that 
diffuse accretion dominates the accretion budget at all redshifts, in each mass bin for both CDM and WDM.
Comparing low-mass and high-mass results for diffuse accretion, in CDM we see that low-mass haloes have lower \Mdif\ than high-mass haloes, while the opposite is true for WDM (see also figure~\ref{Mrates-distri}). For accretion by mergers, on the other hand, high-mass haloes have higher \Mmer\ than low-mass haloes in both CDM and WDM.

And, interestingly, while the high-mass total accretion rates in CDM and WDM are nearly identical, the relative contribution to this total from mergers versus diffuse accretion is higher in CDM than in WDM.

These results are all consistent with an overall picture in which the arrested growth of small-scale structure in WDM inhibits mass accretion through mergers, as compared to CDM. The fact that differences between WDM and CDM accretion rates are dramatic at low masses is not surprising, considering that our low-mass bin is at slightly smaller mass than the half-mode mass for our WDM cosmology (see section~\ref{sec:analytical} for analytical insights into this behaviour). The fact that \emph{high-mass} haloes also show differences between WDM and CDM can be attributed to the difference in available substructure for such haloes in the two cosmologies. In the following sections, we explore the relation between these accretion rates and the local environment of haloes.

\subsection{Environments of haloes accreting with and without mergers}
\label{subsec:medianenv}

\begin{figure*}
\centering
\includegraphics[width=0.85\textwidth,trim= 2 2 10 20,clip]{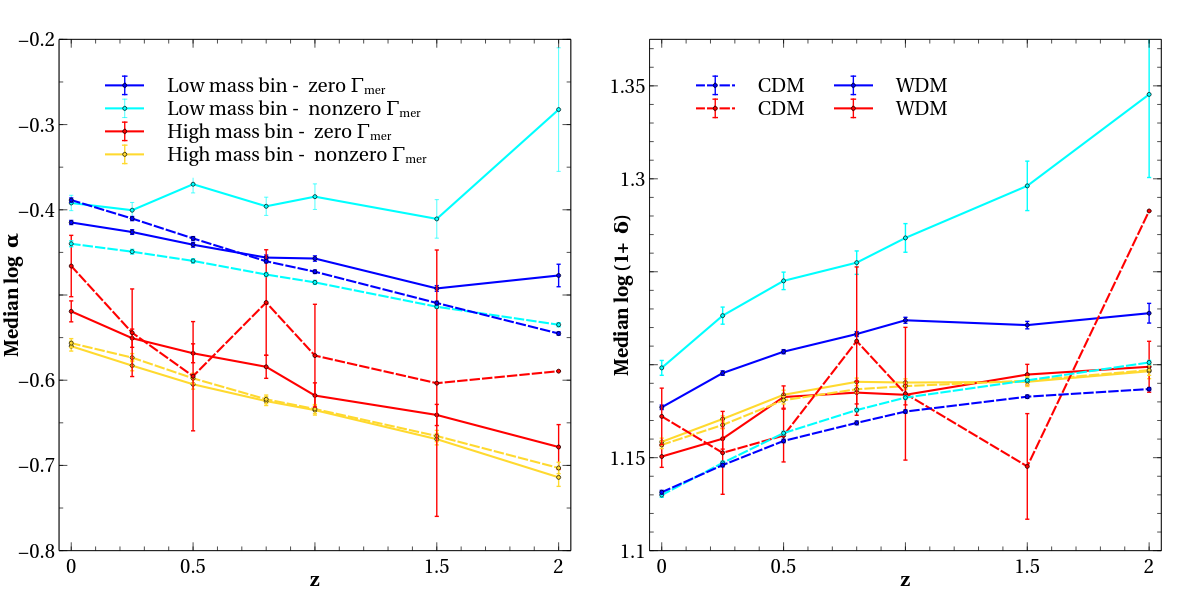}
\caption{{\bf Local environments with and without mergers:} Evolution of median local tidal anisotropy $\alpha$ \emph{(left panel)} and local overdensity $\delta$ \emph{(right panel)} of haloes selected to have grown without mergers (blue, red) and with mergers (cyan, orange) in the previous 2 dynamical times, for the low-mass (cooler colours) and high-mass (warmer colours) bin. Results for CDM (WDM) are shown using dashed (solid) curves. Error bars were  calculated by bootstrap sampling at each redshift.
\underline{\emph{Highlights:}}  While most of these trends conform to expectations  based on the preponderance of small-scale bound structures in CDM as compared to WDM, two trends  are noteworthy: (i) low-mass CDM haloes growing without mergers are in slightly more \emph{anisotropic} environments compared to low mass CDM haloes growing with mergers and (ii) low-mass WDM haloes are in  denser environments than other categories (see also figure~\ref{cdmwdm-delta-4z}). 
} 
\label{median-evoln}
\end{figure*}

To start with, we focus on whether the environmental variables $\alpha$ and $\delta$ evolve differently on average for haloes that add mass with and without mergers, in both CDM and WDM cosmologies. 
Figure \ref{median-evoln} shows the median $\alpha$ \emph{(left panel)} and $\delta$ \emph{(right panel)} for haloes having zero and non-zero merger rates in the low-mass (blue, cyan) and high-mass (red, orange) bins. The solid (dashed) lines show results for WDM (CDM). Error bars on the measurements were  calculated by bootstrap sampling at each redshift. Several interesting trends are apparent, as we discuss next.

Overall, we see that the tidal environments of objects at any fixed mass are typically more anisotropic at later times, while most local densities evolve moderately or decrease at later times. This is broadly consistent with the growth of the cosmic web such that a larger fraction of haloes at fixed mass find themselves in filamentary or sheet-like environments at late times.

Comparing the trends for low-mass and high-mass objects, we see for CDM that high-mass objects (dashed, warm colours) are in relatively denser and more isotropic environments than low-mass objects (dashed, cool colours), at any redshift. This is fully consistent with the standard hierarchical picture in which massive haloes are more clustered and dominate their tidal environments more than low-mass haloes. 

In WDM, while massive haloes do live in more isotropic environments than their low-mass counterparts (solid warm versus solid cool in left panel), unlike CDM, now the \emph{low-mass} haloes are in  denser environments than high-mass ones (same in right panel, see also figure~\ref{cdmwdm-delta-4z}). These low-mass WDM environments are also more overdense than the corresponding low-mass CDM environments.\footnote{We have also checked that the median $b_1$ values for various halo categories are nearly identical between CDM and WDM, and conform to the usual expectation of low-mass objects being less biased than high-mass ones.} We return to this point later.

We focus next on the \emph{high-mass bin} where one might expect similarities between CDM and WDM haloes (although see section~\ref{subsec:mergvsdiff}). Indeed, we see that haloes with mergers occupy nearly identical median density and tidal environments at all redshifts in CDM and WDM (compare the dashed versus solid orange curves in each panel). And, within errors, this is also true for haloes without recent mergers (red curves).

These high-mass environments tend to be the most isotropic of all (c.f. figure~\ref{cdmwdm-alpha-4z}) while having intermediate densities, at almost any redshift.
Thus, most of the differences between high-mass CDM and WDM haloes are related to the lack of substructure and corresponding dominance of diffuse accretion in WDM seen in figure~\ref{MAH}. 

In the \emph{low-mass bin}, the results are more nuanced, with several differences between zero and non-zero merger environments in CDM and WDM which can be summarized as follows:
\begin{itemize}
    \item The environments of low-mass CDM haloes with and without mergers have similar densities at all redshifts (dashed cyan versus dashed blue in the right panel), but are systematically more anisotropic  at low redshift for haloes without mergers at any redshift (same in the left panel). We discuss this further below. 
    \item In contrast, low-mass WDM haloes without mergers (solid blue) live in systematically less dense and more \emph{isotropic} environments than those with mergers (solid cyan). This is sensible, since isolated, isotropic environments would allow diffuse accretion to dominate over mergers. 
    \item Low-mass WDM haloes without mergers (solid blue) live in more dense environments (with similar tidal anisotropy) than their CDM counterparts (dashed blue).
    This is consistent with an enhancement of diffuse accretion in WDM; the lack of substructure in WDM compared to CDM means that environments conducive to purely diffuse accretion (i.e., no recent mergers) can be denser in WDM than in CDM. 
    \item In contrast, low-mass WDM haloes with mergers (solid cyan) live in substantially more dense and more \emph{anisotropic} environments than their CDM counterparts (dashed cyan). We discuss this below. 
\end{itemize}

These low-mass results can be mostly understood keeping in mind the preponderance of small objects in WDM as compared to CDM, as well as the facts that a higher local tidal anisotropy would typically inhibit the accretion of diffuse mass \citep[e.g., by redirecting local flows towards nodes; see][]{bprg17}, while higher densities would enhance mergers. Some of these trends, however, deserve more careful consideration. For example,
the CDM haloes without mergers are in more \emph{anisotropic} environments than those with mergers, contrary to what is suggested by the argument above. Similarly, the higher anisotropy of low-mass WDM haloes with mergers as compared to their CDM counterparts is intriguing as well.
On the  other hand, the higher \emph{density} of low-mass WDM haloes with mergers as compared to all other categories might be understood as being due to the substantially decreased occurrence of mergers in the smooth WDM cosmic web, with only the highest density environments being capable of sustaining merger events.
In the next section, we will explore in more detail the correlations between environment and mass accretion rate, which will shed further light on the nature and origin of these trends.

\subsection{Correlation between mass accretion rates and local halo environment}
\label{subsec:mdot<->env}

To quantify the dependence of mass accretion rates on environment, in this section we study the evolving correlations between our environmental proxies $\alpha$ and $\delta$, and the mass accretion rates due to diffuse accretion (\Mdif) and mergers (\Mmer).

Following \citet{rphs19}, we calculate Spearman rank correlation coefficients between the environmental variables and accretion rates; namely, in each mass bin and for every redshift, we calculate the coefficients $\alpha\leftrightarrow\Mdif$, $\delta\leftrightarrow\Mdif$, $\alpha\leftrightarrow\Mmer$, $\delta\leftrightarrow\Mmer$, as well as the coefficient $\alpha\leftrightarrow\delta$ for reference \citep[which was also studied by][for $z=0$ CDM haloes]{rphs19}. In order to assess which, if any, of the two environmental proxies is more important, we also compute \emph{conditional coefficients}. For stochastic variables $a$, $b$ and $c$, the coefficient between $a$ and $b$, conditioned on $c$, is given by 
\be
(a\leftrightarrow b)|c = (a\leftrightarrow b) - (a\leftrightarrow c)(b\leftrightarrow c)\,. 
\label{eq:conditional-cc}
\ee
If the variables were Gaussian distributed with a joint distribution that had the structure $p(a,b,c) = p(a|c)p(b|c)p(c)$, then the variables $a$ and $b$ would be conditionally independent and we would have $(a\leftrightarrow b)|c = 0$ although $(a\leftrightarrow b)$ need not be zero. We point the reader to \citet{rphs19} for a detailed justification for constructing conditional coefficients using \eqn{eq:conditional-cc} with Spearman rank coefficients, even for variables that are \emph{not} Gaussian distributed, as is the case here (see also  below).

\begin{figure*} 
\centering
\includegraphics[width=0.85\textwidth,trim= 2 2 30 20,clip]{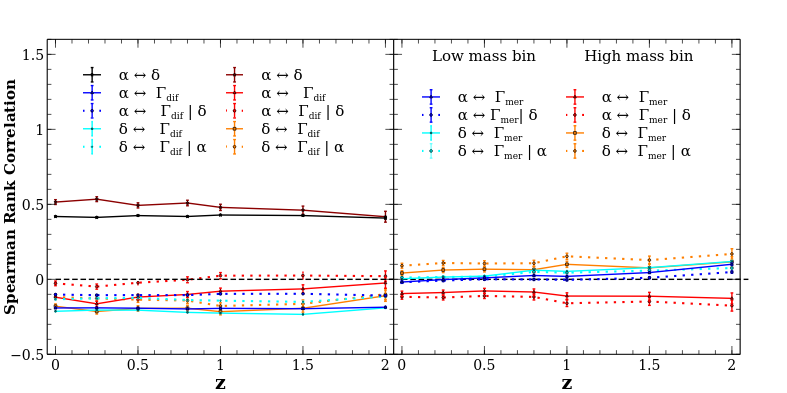}
\caption{{\bf Correlations between accretion rates and local environment (CDM haloes):} Spearman rank correlation coefficients between local environmental variables $\alpha$, $\delta$ and specific accretion rates \Mdif\  \emph{(left panel)} and \Mmer\  \emph{(right panel)}, as a function of redshift. In each panel, cool (warm) colours indicate results for low-mass (high-mass) haloes. Solid curves show the correlations $(\Gamma\leftrightarrow\alpha)$ (red, blue) and $(\Gamma\leftrightarrow\delta)$ (orange, cyan) and, in the \emph{left panel}, $(\alpha\leftrightarrow\delta)$ (brown, black). Dotted curves show conditional coefficients $(\Gamma\leftrightarrow\alpha)|\delta$ and $(\Gamma\leftrightarrow\delta)|\alpha$, calculated using \eqn{eq:conditional-cc}.  
\underline{\emph{Highlights:}} 
(i) There are significant anti-correlations $(\Mdif\leftrightarrow\delta)$ and $(\Mdif\leftrightarrow\alpha)$ in both mass bins (left panel, solid curves). 
(ii) \Mmer\ correlates positively (negatively) with $\delta$ ($\alpha$) in the high-mass bin (right panel, solid orange and red), and does not correlate with either in the low-mass bin (right  panel, solid cyan and blue).
(iii) The conditional coefficient $(\Mdif\leftrightarrow\alpha)|\delta$ for high-mass haloes (left panel, dotted red) almost vanishes at all $z$, indicating that $\delta$ almost fully explains the environment dependence of diffuse accretion in high-mass CDM haloes. 
} 
\label{CDM-spear}
\end{figure*}

\begin{figure*} 
\centering
\includegraphics[width=0.85\textwidth,trim= 2 2 30 20,clip]{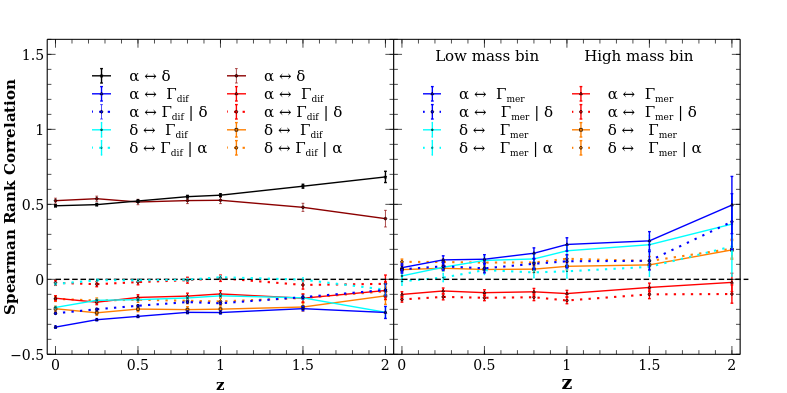}
\caption{Same as figure \ref{CDM-spear}, for {\bf WDM haloes}. 
\underline{\emph{Highlights:}}  (i) The coefficient $(\Mmer\leftrightarrow\alpha)$ for low-mass haloes is, surprisingly, positive (right panel, solid blue) and the corresponding conditional coefficient $(\Mmer\leftrightarrow\delta)|\alpha$ (right panel, dotted cyan) is close to zero. 
(ii) The conditional coefficient $(\Mdif\leftrightarrow\delta)|\alpha$ for low-mass haloes nearly vanishes (left panel, dotted cyan), even though the primary correlation $(\Mdif\leftrightarrow\delta)$ (solid cyan) is significantly non-zero. 
The same is true for the conditional coefficient $(\Mdif\leftrightarrow\alpha)|\delta$ for high-mass haloes (left panel, dotted red).
Thus, $\alpha$ ($\delta$) almost completely explains the environment dependence of diffuse accretion for low-mass (high-mass) WDM haloes. 
} 
\label{WDM-spear}
\end{figure*}

Figure \ref{CDM-spear} (figure \ref{WDM-spear}) shows the correlation coefficients for CDM (WDM). In each case, in the \emph{left (right) panel} we display correlations between the environment and \Mdif\ (\Mmer) as a function of redshift.\footnote{For this correlation analysis, we only evaluate \Mmer\ using those haloes that had mergers in the previous 2 dynamical times, i.e., objects having $\Mmer\neq0$. \Mdif\ on the other hand, is evaluated using all haloes in the bin.} Additionally, the \emph{left panel} in each case also shows the evolving correlation $(\alpha\leftrightarrow\delta)$. Results for the low-mass (high-mass) bin are displayed using cool (warm) colours. The solid curves show the primary correlations such as $(\alpha\leftrightarrow\Mmer)$ while the dotted curves show conditional coefficients such as $(\alpha\leftrightarrow\Mmer)|\delta$. Error bars on the measurements were  calculated by bootstrap sampling at each redshift.  This represents the first detailed study in the literature, of the evolving correlation between local environment and mass accretion rates in either CDM or WDM cosmologies.

We see that $\alpha$ and $\delta$ are always positively correlated for both CDM and WDM, with a strength that is relatively independent of halo mass and that only weakly depends on redshift and dark matter type. This extends the results of \citet{rphs19} (who studied CDM haloes at $z=0$) to significantly higher redshift, as well as to WDM cosmologies. The remaining correlations, which depend only weakly on redshift, can be summarized as follows.

\begin{itemize}
    \item \underline{\emph{High-mass bin:}} 
    \begin{itemize}
        \item 
        For both CDM and WDM, we see a strong negative correlation $(\delta\leftrightarrow\Mdif)$, and a weaker but significant positive correlation  $(\delta\leftrightarrow\Mmer)$, at all redshifts (solid orange). This is consistent with the expectation that overdense environments enhance mergers while underdense regions allow for more diffuse accretion \citep{fm10}. 
        
        \item The corresponding primary correlations $(\alpha\leftrightarrow\Mdif)$ and $(\alpha\leftrightarrow\Mmer)$ (solid red) are all negative, with magnitudes comparable to those in the case of $\delta$, for both CDM and WDM. Interestingly, however, the \emph{conditional} coefficient $(\alpha\leftrightarrow\Mdif)|\delta$ (dotted red  in left panels) for  both CDM and WDM almost vanishes at all $z$, while  the corresponding conditional coefficient $(\alpha\leftrightarrow\Mmer)|\delta$ does not vanish (dotted red in right panels).\footnote{When using a mass threshold $M_{\rm vir}>4.84\times10^{11}\Mh$ for defining progenitors (see section~\ref{subsec:Mdot-measure}), the high-mass negative $(\alpha\leftrightarrow\Mmer)$ correlation becomes weaker, and in fact slightly positive at low redshift. This is the only qualitative difference we find between different progenitor definitions.} \emph{Thus, for high-mass haloes, $\alpha$ and $\delta$ are equally important for accretion via mergers, while $\delta$ almost completely explains the environment dependence of diffuse accretion.}
        
    \end{itemize}
    \item \underline{\emph{Low-mass bin:}}
    \begin{itemize}
        \item In low-mass CDM haloes, \Mmer\ shows almost no correlation with either $\alpha$ or $\delta$ at any redshift, consistent with similar weak correlations seen previously using other environmental proxies \citep[c.f. figure 4 of][]{fm10}.
        
        \item The situation changes in WDM: we now see \emph{positive} correlations $(\alpha\leftrightarrow\Mmer)$ and $(\delta\leftrightarrow\Mmer)$, comparable in strength to each other and increasing at higher redshift. Moreover, at least at low redshift, we see that the conditional coefficient $(\delta\leftrightarrow\Mmer)|\alpha$ (dotted cyan) is close to zero. In fact, this is the case within errors at all $z\lesssim1.5$. \emph{Thus, $\alpha$ mostly accounts for the environment dependence of \Mmer\ at nearly all redshifts for low-mass WDM haloes, with \Mmer\ being enhanced in more anisotropic tidal environments.}

        \item For \Mdif\ in low-mass CDM haloes, both $(\delta\leftrightarrow\Mdif)$ and $(\alpha\leftrightarrow\Mdif)$ are negative, with comparable magnitudes (solid cyan and blue). Also, neither of the conditional coefficients $(\alpha\leftrightarrow\Mdif)|\delta$ and $(\delta\leftrightarrow\Mdif)|\alpha$ is close to zero (dotted cyan and blue). Thus, both $\alpha$ and $\delta$ play a role for \Mdif, consistent with the expectation that diffuse accretion should be more efficient far from crowded regions, i.e. in underdense, isotropic environments in CDM.
        
        \item For \Mdif\ in low-mass WDM haloes, on the other hand, we see a stronger $(\alpha\leftrightarrow\Mdif)$ anticorrelation (solid blue), which almost completely explains the $(\delta\leftrightarrow\Mdif)$ (solid cyan) anticorrelation, seen as the near-vanishing  of $(\delta\leftrightarrow\Mdif)|\alpha$ (dotted cyan) at  all redshifts. \emph{Thus, the environment dependence of \Mdif\ in low-mass haloes is related to both $\alpha$ and $\delta$ for CDM and almost completely explained by $\alpha$ for WDM.}
        The dominant role played by the tidal environment for low-mass WDM haloes is consistent with previous simulation results \citep{angulo} which indicate that these objects form in a well-established tidal environment which can then further affect their mass accretion.
        
    \end{itemize}
\end{itemize}

The nature of the environmental correlations discussed above reveals a complex interplay between mass accretion rates and local environment. While this is perhaps not surprising, considering the intimate connection between mass accretion in the halo outskirts and the internal structure of haloes \citep[e.g.,][]{dk14,mdk15}, our results above show several interesting aspects. In section~\ref{subsec:medianenv}, for example, we saw that low-mass WDM haloes with mergers live in systematically more anisotropic environments than their CDM counterparts, a trend consistent with the correlations seen above in which $(\alpha\leftrightarrow\Mmer)$ is positive for WDM but negative for CDM. This is counter-intuitive, considering that accretion due to mergers is enhanced in overdense, isotropic environments in CDM, and one might expect this trend to be even more pronounced in WDM which has less small-scale structure available in all environments. 
Moreover, the small magnitude of $(\delta\leftrightarrow\Mmer)|\alpha$ for these low-mass WDM haloes shows that $\alpha$, in fact, dominates the environmental trends for \Mmer. This connection between mergers and the local tidal environment of low-mass WDM haloes deserves further study.

Another puzzling result from section~\ref{subsec:medianenv} was that the environments of low-mass CDM haloes without mergers are more anisotropic than of those with mergers. In this case as well, the correlation results above do not particularly clarify the situation, since they are consistent with \Mdif\ being enhanced in underdense, isotropic environments (with both $\delta$ and $\alpha$ playing comparable roles) and \Mmer\ being nearly uncorrelated with environment. We have checked that these trends are not restricted to the median $\alpha$ alone; rather, the entire distributions of $\alpha$ in such haloes are shifted relative to each other.

As a check on systematic errors due to our use of correlation coefficients and \eqn{eq:conditional-cc}, we repeated the analysis by measuring the median accretion rates \Mdif\ and \Mmer\ in narrow bins of $\alpha$ and $\delta$. The results, although noisy, are fully consistent with the conclusions above.

\subsection{Accretion rate assembly bias}
\label{subsec:assemblybias}
In this section, we explore the nature of the correlations between mass accretion rate and the \emph{large-scale} halo environment (characterised by halo bias $b_1$; see section~\ref{subsec:bias}) in fixed mass bins, also known as halo assembly bias or secondary bias. 

Previous studies have largely focused on secondary halo variables such as age, concentration, shape, angular momentum and velocity dispersion structure (see the Introduction for references). To our knowledge, the only previous work that studied the assembly bias of mass accretion rates was by \citet{lazeyras}, who focused on the total mass accretion rates of high-mass CDM haloes at $z=0$. Our analysis below therefore substantially extends these results in terms of halo mass, redshift range and dark matter type.

\begin{figure*} 
\centering
\includegraphics[width=0.85\textwidth,trim= 2 2 20 20,clip]{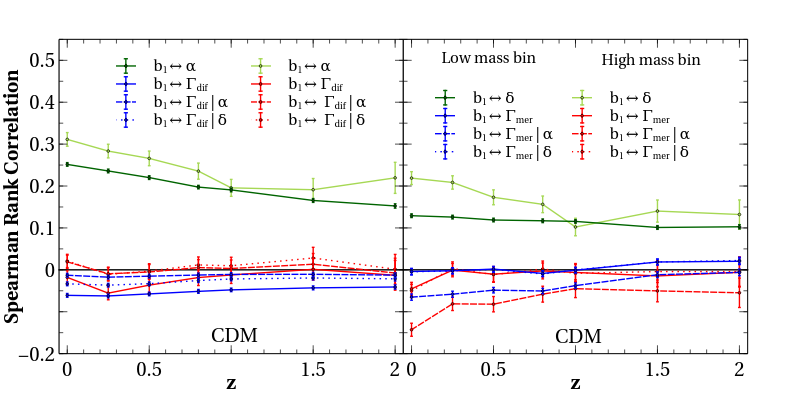}
\caption{
{\bf Accretion rate assembly bias (CDM haloes):} Spearman rank correlation coefficients between large-scale halo bias $b_1$ and specific accretion rates \Mdif\  \emph{(left panel)} and \Mmer\  \emph{(right panel)} for low-mass (blue) and high-mass (red) CDM haloes, as a function of redshift. Solid curves show the correlations $(\Gamma\leftrightarrow b_1)$ and, in the \emph{left [right] panel}, green curves show $(b_1\leftrightarrow\alpha)$ [$(b_1\leftrightarrow\delta)$], with light (dark) green for high-mass (low-mass) haloes. Dashed and dotted curves respectively show conditional coefficients $(\Gamma\leftrightarrow b_1)|\alpha$ and $(\Gamma\leftrightarrow b_1)|\delta$, calculated using \eqn{eq:conditional-cc}. Error bars were estimated using bootstrap sampling at each redshift. \underline{\emph{Highlights:}} There is a significant negative correlation $(\Mdif\leftrightarrow b_1)$ for low-mass haloes at all redshifts (left panel, solid blue), which is  largely explained by the $(\Mdif\leftrightarrow\alpha)$ correlation (conditional coefficient shown by the dashed blue curve is close to zero). 
} 
\label{CDM-spearbias}
\end{figure*}

\begin{figure*} 
\centering
\includegraphics[width=0.85\textwidth,trim= 2 2 20 20,clip]{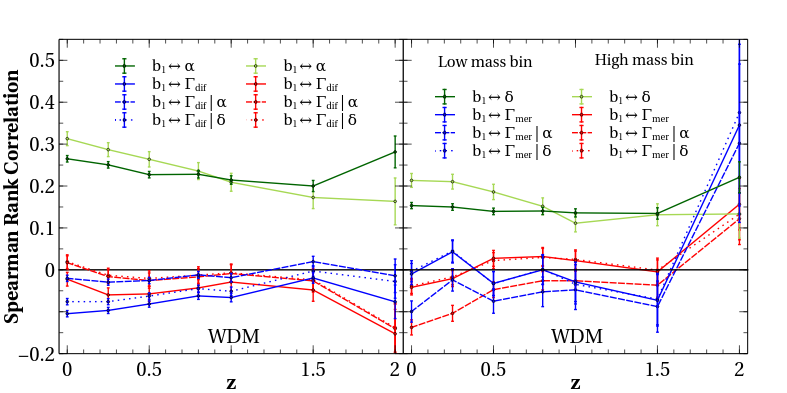}
\caption{Same as figure \ref{CDM-spearbias}, for {\bf WDM haloes}. \underline{\emph{Highlights:}} There is a significant negative correlation $(\Mdif\leftrightarrow b_1)$ for low-mass haloes  (left panel, solid blue), which increases in magnitude at low redshift, is stronger than the corresponding correlation for CDM haloes and is mostly explained by the $(\Mdif\leftrightarrow\alpha)$ correlation (conditional coefficient shown by the dashed blue curve is close to zero).
} 
\label{WDM-spearbias}
\end{figure*}

We define the assembly bias of mass accretion as the two correlations $(b_1\leftrightarrow\Mdif)$ and $(b_1\leftrightarrow\Mmer)$ measured using Spearman rank correlation coefficients in different mass bins for all redshifts. The results of \cite{rphs19} indicate that, at $z=0$ in CDM, the assembly bias of each of the internal properties halo concentration, spin, shape and velocity dispersion can be mostly attributed to two fundamental correlations, one between $b_1$ and $\alpha$ and the other between $\alpha$ and the halo internal property. Considering that there are relatively strong correlations between the mass accretion rates \Mmer, \Mdif\ and the local environmental proxies $\alpha$ and $\delta$ (see figures~\ref{CDM-spear} and~\ref{WDM-spear}), it is interesting to ask (a) what is the overall strength of assembly bias with mass accretion rate in comparison to other internal halo properties, and (b) what role, if any, do the local environmental proxies play in explaining these trends?

Figure \ref{CDM-spearbias} (figure \ref{WDM-spearbias}) shows the results for CDM (WDM) haloes. As with figures~\ref{CDM-spear} and~\ref{WDM-spear}, the \emph{left (right) panels} show results for \Mdif\ (\Mmer), with the solid (dotted) curves showing primary (conditional) coefficients. Additionally, the \emph{left (right) panels} show the correlation $(b_1\leftrightarrow\alpha)$ ($(b_1\leftrightarrow\delta)$) for comparison.

Similarly to the results for $(\alpha\leftrightarrow\delta)$ in figures~\ref{CDM-spear} and~\ref{WDM-spear}, we see that the correlations $(b_1\leftrightarrow\alpha)$ and $(b_1\leftrightarrow\delta)$ are significantly positive, very similar across halo mass as well as between CDM and WDM, and evolve moderately between $z=2$ and $z=0$. The $(b_1\leftrightarrow\alpha)$ correlation at any redshift is $\sim50\%$ larger than the corresponding $(b_1\leftrightarrow\delta)$ correlation, for both CDM and WDM. This extends the $z=0$ CDM results of \citet{phs18a} and \citet{rphs19} to significantly higher redshifts and to WDM cosmologies.

In the \emph{high-mass bin}, the correlations $(b_1\leftrightarrow\Mdif)$ and $(b_1\leftrightarrow\Mmer)$ are relatively weak at nearly all redshifts for both CDM and WDM. The exception is at the highest redshift $z=2$ for WDM, where $(b_1\leftrightarrow\Mdif)$ is  negative while $(b_1\leftrightarrow\Mmer)$ is  positive, but with large errors. Overall, therefore, we conclude that high-mass objects in CDM and WDM do not show significant assembly bias for either mass accretion rate.

In the \emph{low-mass bin}, we see more interesting results. Although the $(b_1\leftrightarrow\Mmer)$ correlation is essentially zero across all redshifts for both CDM and WDM, the same is not true for the $(b_1\leftrightarrow\Mdif)$ correlation, which remains significantly negative at nearly all redshifts, in both CDM and WDM. In fact, the strength of the correlation at $z=0$, namely $|(b_1\leftrightarrow\Mdif)|\gtrsim 0.05\, (0.1)$ for CDM (WDM), is large and comparable to the strongest correlations seen for any secondary variable, at this halo mass. \citep[see, e.g., in the middle panel of figure 2 in ][]{rphs19}. If we extrapolate the results in \citep{rphs19} to the low masses $(1.2 \times 10^{11} \Mh < M_{vir} < 3.85 \times 10^{11} \Mh)$ we study in this paper, we find $b_1 \leftrightarrow c$ correlation to be  $ \sim 0.07(0.09)$ for CDM(WDM).
The magnitude of the correlation is nearly constant with redshift for CDM while showing an increasing trend with time for WDM.
\emph{Thus, the assembly bias of low-mass accretion rates, especially at low redshift, is a comparatively large effect, entirely driven by  diffuse accretion.} 

More interestingly, the conditional coefficient $(b_1\leftrightarrow\Mdif)|\alpha$ for  low-mass haloes is close to zero at all redshifts, for both CDM and WDM. (For CDM, the coefficient $(b_1\leftrightarrow\Mdif)|\delta$ also has a small magnitude, but is not as close to zero as $(b_1\leftrightarrow\Mdif)|\alpha$.) 

\emph{Thus, the local tidal anisotropy plays a key role in explaining the accretion rate assembly bias of both CDM and WDM haloes}, with local overdensity also playing a significant role in CDM. 
These findings are consistent with previous results in the literature \citep[e.g.,][]{dalal, hpdc09,bprg17} which show that low-mass haloes which are in the vicinity of massive haloes,  and are hence highly clustered at large-scales while residing in anisotropic local tidal environments, show suppressed (diffuse) accretion rates.

\section{Analytical insights}
\label{sec:analytical}
In this section, we discuss the numerical results presented above in the language of the excursion set approach \citep{bcek91,lc93,lc94,bm96}. Below, we briefly recapitulate the basic concepts underlying the excursion set approach and its modern variants, before discussing our results in this context.

\subsection{Recap of the excursion  set (peaks) approach}

In the excursion set framework, halo abundances, large-scale clustering and accretion rates are estimated by identifying and counting likely locations of virialisation identified in the \emph{initial} stochastic density field (linearly extrapolated to present epoch). A key ingredient in the traditional excursion set approach is a choice of collapse threshold or `barrier' $B$ which must be crossed by random walks in the linearly extrapolated density field $\delta_R$ as a function of Lagrangian smoothing scale $R$. The barrier $B$ is usually adopted from spherical \citep{gg72} or ellipsoidal \citep{smt01} collapse models, and depends on the redshift $z$ of interest and possibly additional stochastic variables \citep[such as those related to tidal effects in the ellipsoidal model, see][]{cphs17}.

The statistical properties of the random walks, on the other hand, are determined by the choice of initial matter power spectrum and a smoothing filter, which fixes the functional relation $\sigma^2(m)$ where $\sigma^2$ is the variance of the linearly extrapolated density contrast smoothed on a Lagrangian scale corresponding to mass $m\sim R^3$. 
If a random walk first upcrosses the barrier $B(z)$ at scale $R(m)$ (starting from $R\to\infty$ or $\sigma^2\to0$), then a halo of mass $m$ is declared to form at redshift $z$.
While early work on the subject focused on a filter that is sharp in Fourier space (for reasons of analytical simplicity), later developments have shown how to efficiently analyze the effects of more realistic filters that are compact in real space and lead to  random walks with correlated steps as $R$ is varied (\citealp{ms12}; see also \citealp{bcek91,zentner07,pls12}). Finally, the recognition that the sites of virialisation are special \citep{smt01} -- e.g., peaks in the linearly extrapolated density field \citep{bbks86} -- leads to an `excursion set peaks' approach \citep[ESP,][]{ps12b,aj90} which has been shown to agree with simulated halo mass functions and large-scale clustering in CDM cosmologies at the $\sim10\%$ level \citep{psd13,cphs17}. 

The ESP framework has also been studied in the context of WDM cosmologies by \citet[][hereon, HP14]{hp14}, who demonstrated that the suppression of small-scale power in these models is an excellent diagnostic tool for testing the assumptions underlying the excursion set (peaks) approach. In particular, HP14 argued that a single-barrier framework is insufficient to properly explain halo abundances due to small but systematic errors in the ellipsoidal collapse model \citep{Monaco99,gmst07,lbp14}, which become dramatically amplified in WDM models as compared to CDM. Here, however, we are interested in the prediction \citep{lc93,lc94} that mass accretion rates are intimately connected with the slope of the initial power spectrum near the scales of interest, with steeper spectra leading to higher accretion rates. The consequences of the latter effect for WDM can be understood without delving  into the details of any specific excursion set model and only depend on the correlated nature of the random walks, as we discuss next.

\subsection{Accretion rates in the ESP framework}

In the following, in addition to the barrier $B$ for linearly extrapolated density fluctuations and the $\sigma(m)$ relation mentioned above, we will need two more quantities from the excursion set lexicon. The first is the proto-halo `significance' $\nu(m,z) = \delta_{\rm c}(z)/\sigma(m)$, where $\delta_{\rm c}(z) \propto 1/D(z)$ is the collapse threshold from the  spherical model, with $D(z)$ being the linear theory growth factor. The second is the typical peak curvature $\avg{x|\nu}$ at scale $\nu(m,z)$, where $x=-\nabla^2\delta_R/\sqrt{{\rm Var}(\nabla^2\delta_R)}$ \citep[see figure 6 of][]{bbks86}.

As mentioned above, accretion rates in CDM and WDM are expected to be very different due to the comparative steepness of the \emph{initial} matter power spectrum in WDM around the half-mode mass scale. To see what this implies, we must mainly keep in mind that the $\sigma(m)$ relation becomes `stretched out' for a truncated power spectrum such as WDM, with $\sigma(m)\to$ a constant below the half-mode mass (see, e.g., figure 2 of HP14). In CDM, on the other hand, $\sigma(m)$ continues to increase down to very small masses. A simple calculation now shows that the diffuse mass accretion rate can be written as \citep[e.g.,][]{lazeyras}

\begin{align}
    \Gamma = \frac{\der\ln M}{\der\ln a} &= \frac{1}{\sigma}\,\frac{1}{\der B/\der\sigma} \frac{\der B/\der\ln a}{\der\ln\sigma/\der\ln M} \notag\\
    &\approx \frac{1}{\sigma}\,\frac{\gamma}{\avg{x|\nu}}\,\frac{|\der B/\der\ln a|}{|\der\ln\sigma/\der\ln M|}\,,
    \label{eq:hp14-accrate}
\end{align}
where the first line is a manipulation of variables and the second line relates the slope at barrier crossing, $\der B/\der \sigma$, to the peak curvature $\avg{x|\nu}$ \citep{ms12,ps12b}, with $\gamma$ being a spectral variable of order unity in both CDM and WDM ($\gamma\simeq0.5$ for our cosmology).

At high masses, for both CDM and WDM, $\avg{x|\nu}$ is large, $\sigma$ is small and $|\der\ln\sigma/\der\ln M|$ is finite. At masses smaller than the half-mode mass for WDM, $\avg{x|\nu}$ is smaller than at high masses, $\sigma$ is \emph{constant} and $|\der\ln\sigma/\der\ln M|$ approaches zero. This last feature makes the accretion rates very high. At low masses for CDM, on the other hand, $\avg{x|\nu}$ is similar to that in WDM, $\sigma$ is significantly \emph{larger} than at high masses and $|\der\ln\sigma/\der\ln M|$ is finite, so that low-mass CDM accretion rates remain small. This qualitatively explains the overall trends seen in the right panel of figure~\ref{MAH}.

Quantitatively, in our high-mass bin at $z=0$, for both CDM and WDM we have $\sigma\lesssim1.5$,  $\avg{x|\nu}\gtrsim3$ and $|\der\ln\sigma/\der\ln M|\lesssim0.2$. Assuming that the time dependence of the barrier is completely determined by $\delta_{\rm c}(z)$, we also have $|\der B/\der\ln a|\sim1$ at $z=0$. Equation~\eqref{eq:hp14-accrate} then predicts high-mass accretion rates of $\Gamma\sim0.5$ at $z=0$. 
In the low-mass bin, for CDM, we have $\sigma\sim2.5$, $\avg{x|\nu}\sim2.5$  and $|\der\ln\sigma/\der\ln M|\lesssim0.15$. For WDM, on the other hand, while $\sigma$ and $\avg{x|\nu}$ are not very different from CDM for our cosmology and choice of mass bin, we have $|\der\ln\sigma/\der\ln M|\lesssim0.075$, a factor 2 smaller than in CDM. Thus low-mass accretion rates at $z=0$ are predicted to be about a factor 2 higher for our WDM low-mass bin than the corresponding CDM values, which agrees with the trend seen in the right panel of figure~\ref{MAH}. The values predicted, $\Gamma\sim0.5\,(1.0)$ for CDM (WDM), are higher than the measured ones, which is perhaps not surprising considering the approximate nature of our  calculation.\footnote{We have also ignored the issue of mass reassignment discussed by HP14, in which the $\sigma(m)$ relation must effectively be modified for collapsed objects, with the effect becoming more prominent in WDM at scales substantially smaller than the half-mode mass.}

The discussion of environmental trends for mass accretion rates requires a prediction of the cross-correlation between the slopes of random walks at barrier crossing and the values attained by these walks at larger smoothing scales. This is conceptually easiest for the density environment, which is the natural variable used in excursion set calculations. \citet{lazeyras} have applied such arguments to show that an overall assembly bias trend, in which slowly accreting haloes are strongly clustered as compared  to rapid accretors, i.e. a negative correlation $(b_1\leftrightarrow\Gamma)$, is a natural prediction of the excursion set peaks framework. The distinction between diffuse accretion and that via mergers, on the other hand, requires a higher level of sophistication in the simultaneous prediction of continuous and discrete accretion rates, an aspect which excursion set models have only recently begun to explore \citep{ms14-markov}. And the trends we have noted with tidal anisotropy $\alpha$, especially the result that $\alpha$ largely explains low-mass assembly bias trends with mass accretion,  are currently not predictable by any excursion set (peaks) model that  we are aware of \citep[although see][for some initial steps in this direction]{cphs17}. We therefore defer a discussion of analytical predictions for the environment dependence of mass accretion to future work, where we hope to develop a versatile excursion set framework that can address these issues.

\section{Summary \& Conclusions}
\label{sec:conclude}

We have investigated the evolving correlations between mass accretion rates and environment in $N$-body simulations of CDM and WDM cosmologies. The latter are characterised by a strong suppression of small-scale power (we deliberately chose a somewhat extreme case with $m_{\rm dm}=0.4\,{\rm keV}$), which creates dramatic differences in the nature of mass accretion between haloes smaller and larger than the half-mode mass scale. 

While accretion rates in WDM models have been compared with their CDM counterparts previously in the literature \citep{knebe02, benson+13, elahi, khimey20}, and environmental trends of CDM accretion rates have also been studied before (see the Introduction), our analysis represents the first systematic comparison of environmental trends of accretion rates between CDM and WDM haloes, at masses above and below the WDM half-mode mass, using multiple proxies for the local as well as large-scale environment, and over a wide range of redshift $2\geq z\geq 0$.

Most of our results (section~\ref{sec:results}) regarding the evolving nature and environment dependence of specific accretion rates due to mergers (\Mmer) and diffuse mass (\Mdif; see equations~\ref{eq:Mdot-def}) can be understood in terms of the lack of small-scale structure below the half-mode mass scale in WDM, which affects not only the low-mass haloes but also the accretion histories of massive objects. We summarize these below and also highlight a few puzzling findings that deserve further study.

\vskip 0.1in
\noindent
\underline{\it Environment-independent trends:}
\begin{itemize}
    \item Specific accretion rates of low-mass CDM haloes are lower than those of high-mass CDM haloes, while low-mass WDM haloes have  higher accretion rates than their high-mass counterparts (figure~\ref{MAH}). This is a straightforward prediction of the  excursion set approach with correlated steps (section~\ref{sec:analytical}). 
    
    \item Mass accretion in WDM haloes is dominated by \Mdif, consistent with previous work \citep{benson+13,elahi}. \Mdif\ (\Mmer) in WDM haloes is also higher (lower) than that in CDM haloes of the same mass,  as expected from the lack of small-scale bound structures in WDM. 
\end{itemize}

\noindent
\underline{\it Trends with local environment:}
\begin{itemize}
    \item The evolving median local density and tidal anisotropy of haloes with and without mergers (figure~\ref{median-evoln}) are also largely consistent with expectations based on the lack of small-scale structure in WDM. The only exceptions, which deserve further study, are
    \begin{itemize}
        \item the environments of low-mass WDM haloes are denser than those of all other categories, and
        \item the environments of low-mass CDM haloes without mergers are more anisotropic than of those with mergers. 
        
    \end{itemize}
    \item The correlations between environment and accretion rates in  figures~\ref{CDM-spear} and~\ref{WDM-spear} add further detail to these results. For both CDM and WDM,  at all redshifts in the \emph{high-mass bin},
        \begin{itemize}
            \item there is a strong negative correlation between $\delta$ and \Mdif, a weaker positive correlation between $\delta$ and \Mmer, and comparable negative correlations between $\alpha$ and both \Mdif\ and \Mmer,
            \item while $\alpha$ and $\delta$ are equally important for \Mmer, $\delta$ almost completely explains the environment dependence of \Mdif.
    \end{itemize}
    \item In the \emph{low-mass bin} at all $z$, 
    for CDM haloes, $\alpha$ and $\delta$ are equally important in explaining the environment dependence of \Mdif\ and show no correlation with \Mmer. In WDM, on the other hand,
    the environment dependence of both \Mmer\ and \Mdif\ is almost fully explained by $\alpha$. In fact, the correlation $(\alpha\leftrightarrow\Mmer)$ is \emph{positive}, which is counter-intuitive and deserves further study.

\end{itemize}
\emph{In summary, at all $z$, the local tidal anisotropy $\alpha$ plays a completely dominant role for both \Mmer\ and  \Mdif\ in low-mass WDM haloes, and a bigger role than the local density $\delta$ for \Mmer\ in high-mass haloes in both CDM and WDM. In contrast, $\delta$ plays the  dominant role for \Mdif\ in high-mass CDM haloes.
}
In the other cases, $\delta$ and $\alpha$ play equally important roles. These trends are mostly insensitive to the definition of progenitor (see section~\ref{subsec:Mdot-measure} and \ref{subsec:mdot<->env}) for details.

\vskip 0.1in
\noindent
\underline{\it Assembly bias:}
\vskip 0.05in
\noindent
We defined assembly bias using the Spearman rank correlation coefficients $(b_1\leftrightarrow\Mdif)$ and $(b_1\leftrightarrow\Mmer)$ in each mass bin \citep[c.f.,][]{rphs19}.
We find that the following holds for both CDM and WDM haloes (figures~\ref{CDM-spearbias} and~\ref{WDM-spearbias}):
\begin{itemize}
    \item $(b_1\leftrightarrow\alpha)$ is  always higher than $(b_1\leftrightarrow\delta)$ at any $z$. This extends the $z=0$ CDM results of \citet{phs18a} and \citet{rphs19} to higher redshifts and WDM cosmologies.
    \item \emph{We detect assembly bias in the low-mass bin, driven entirely by \Mdif, with a strong $(b_1\leftrightarrow\Mdif)$ correlation at $z=0$ with the strength that is comparable with the highest known correlations of all the secondary variables at this halo mass for CDM and is higher in WDM than in CDM}  \citep[solid blue curves in the left panels of figures~\ref{CDM-spearbias} and~\ref{WDM-spearbias}, see section \ref{subsec:assemblybias} for details and compare the middle panel of figure 2 in][]{rphs19}.  In the high-mass bin, there is no significant assembly bias at any but the highest redshifts we study.
    \item The tidal anisotropy $\alpha$ plays a dominant role in explaining the low-mass assembly bias, especially for WDM haloes.
\end{itemize}

Our results place important constraints on (semi-) analytical excursion set models of halo formation and growth and, by extension, on galaxy evolution models built upon such approximate techniques (see the Introduction for references). As argued by HP14, any such model which purports to explain environmental trends of evolving haloes must logically work equally well for CDM and WDM, since the physics of collisionless self-gravitating systems is common to both. As we have seen, however, the suppression of small-scale power in WDM leads to an intricate dependence of the mass accretion on local (and large-scale) halo environment, particularly at masses smaller than the WDM half-mode mass. As with the mass function at these scales discussed by HP14, producing an accurate model of mass accretion with the correct environment dependence is likely to reveal interesting features of collisionless dynamics in the shell-crossed regime. It will be very interesting to confront our results above with excursion set models of mergers and mass accretion \citep{lc94,mkby11,ms14-markov}, as well as excursion set-inspired semi-analytical algorithms tuned to reproduce CDM results \citep{sk99,pch08,jvdb14}. We leave this to future work.

\section*{Acknowledgements}
We thank Oliver Hahn and Sujatha Ramakrishnan for useful discussions.
PD thanks IUCAA for hospitality and working facilities, Akhilesh Peshwe (Principal, DMPDM Science College) for his kind support, Isha Pahwa for useful discussions and Dhairyashil Jagadale for constant support and discussion.
The research of AP is supported by the Associateship Scheme of ICTP, Trieste and the Ramanujan Fellowship awarded by the Department of Science and Technology, Government of India. 
This work used the open source computing packages NumPy \citep{vanderwalt-numpy},\footnote{\href{http://www.numpy.org}{http://www.numpy.org}} SciPy \citep{scipy}\footnote{\href{http://www.scipy.org}{http://www.scipy.org}} and the plotting software Veusz.\footnote{\href{https://veusz.github.io/}{https://veusz.github.io/}}
We gratefully acknowledge the use of high performance computing facilities at IUCAA, Pune.

\section*{Data availability}
The data underlying this work will be shared upon reasonable request to the authors.

\bibliography{mybib}

\label{lastpage}

\end{document}